\documentclass[twocolumn,tighten]{aastex631}
\usepackage{CJK}
\usepackage{amsmath}
\usepackage{orcidlink}
\usepackage[hyperref]{backref}

\begin{document}

\title{Cosmological Parameter Estimate from Persistent Radio Sources of Fast Radio Bursts }

\begin{CJK*}{UTF8}{gbsn}

\author[0000-0002-9110-4336]{Zi-Liang Zhang (张子良)}\thanks{ziliang.zhang@unlv.edu}
\affiliation{Nevada Center for Astrophysics, University of Nevada, Las Vegas, NV 89154, USA}
\affiliation{Department of Physics and Astronomy, University of Nevada, Las Vegas, NV 89154, USA}

\author[0000-0002-9725-2524]{Bing Zhang (张冰)}\thanks{bing.zhang@unlv.edu}
\affiliation{Nevada Center for Astrophysics, University of Nevada, Las Vegas, NV 89154, USA}
\affiliation{Department of Physics and Astronomy, University of Nevada, Las Vegas, NV 89154, USA}




\begin{abstract}
We introduce a novel method to constrain the Hubble constant (\(H_0\)) by combining fast radio bursts (FRBs) and their persistent radio sources (PRSs) through the observationally validated Yang relation \citep{Yang2020}, $ L_{\nu}  \propto | \mathrm{RM} | $, which links PRS luminosity to the rotation measure (RM) of the associated FRB. Using a mock sample of PRSs, we demonstrate that the Yang relation can help to unravel the degeneracies among \(H_0\), baryon density parameter \(\Omega_b\), and  baryon fraction in the intergalactic medium \(f_{\mathrm{IGM}}\) in the traditional approach of using dispersion measure only to perform cosmological analyses.  Our method employs a two-stage Markov Chain Monte Carlo (MCMC) analysis to constrain \(H_0\). Using the available data of six observed PRS systems, we obtain a preliminary constraint of \(H_0 = 75 \pm 30~\mathrm{km\,s^{-1}\,Mpc^{-1}}\). We briefly discuss possible refinements of the method by reducing residual degeneracies and systematic uncertainties using future data and physical modeling. Our results indicate that the Yang relation can potentially become a new probe for performing FRB cosmology. 
\end{abstract}



\section{Introduction} 
Fast Radio Bursts (FRBs) are millisecond-duration radio transients of extragalactic origin \citep{Lorimer2007,Thornton2013, Cordes2019, Petroff2022, Zhang2023} with the potential of serving as a new type of  cosmological probe \citep{Bhandari2021, Glowacki2024, Wu2024}. 

The dispersion measure (DM), defined as the column density of free electrons along the line of sight, serves as a key diagnostic by encoding information about the electron distribution in intervening  environments. Consequently, FRB-based studies have been employed to trace missing baryons in the intergalactic medium \citep[e.g.,][]{Deng2014,Macquart2020, Li2020}, characterize inhomogeneities in the IGM \citep[e.g.,][]{McQuinn2014}, map large-scale fluctuations \citep[e.g.,][]{Macquart2020, Wu2022}, constrain the Hubble constant \(H_0\) \citep[e.g.,][]{Wu2022}, reconstruct reionization history \citep[e.g.,][]{Deng2014,Zheng2014,Caleb2019,Beniamini2021}, and probe dark energy \citep[e.g.,][]{Zhou2014, Gao2014, Wang2025}.

Currently, there is an inconsistency in measuring the Hubble constant \( H_0 \) using different methods. Measurements of \(H_0\) from late-universe distance-ladder methods (e.g., \(H_0 = 73.04 \pm 1.04~\mathrm{km\,s^{-1}\,Mpc^{-1}}\), \citealt{Riess2022}) conflict with those derived from early-universe cosmic microwave background observations (e.g., \(H_0 = 67.4 \pm 0.5~\mathrm{km\,s^{-1}\,Mpc^{-1}}\),  \citealt{Planck2018}).  Independent measurements of \( H_0 \) would be helpful to resolve the tension, and FRBs offer such an opportunity.

However, measurements of \(H_0\) using the intergalactic DM component, \(\mathrm{DM}_{\mathrm{IGM}}\), are limited by inherent parameter correlations. As shown in Equation~\ref{eq:DMigm}, \(\mathrm{DM}_{\mathrm{IGM}}\) introduces degeneracies among \(H_0\), the baryon density parameter \(\Omega_b\), and the baryon fraction in the IGM, \(f_{\mathrm{IGM}}\)\footnote{The parameter $f_{\rm IGM}$ here denotes the fraction of baryons in the diffuse medium between the FRB host and Milky Way, which strictly also includes contributions from the intervening galaxy halos along the line of sight, so in the literature it is also called $f_{\rm diff}$ \citep{Prochaska2019,zhang2025b}.}.
Further uncertainties arise from intrinsic variations in \(\mathrm{DM}_{\mathrm{IGM}}\) and its degeneracy with host galaxy DM contributions.  

Persistent Radio Sources (PRSs) are steady radio emitters, typically observed at gigahertz frequencies with luminosities of around \(10^{29}\,\mathrm{erg\,s^{-1}\,Hz^{-1}}\), and confined to regions smaller than a parsec. These sources are associated with a subset of repeating FRBs. The first confirmed PRS was discovered in 2017, linked to FRB 20121102A \citep{Chatterjee2017}. Observations of its size, luminosity, and spectral properties rule out the possibility of a coincidental association with nearby star formation activity \citep{Chatterjee2017, Tendulkar2017, Bassa2017, Kokubo2017}, indicating that PRSs are associated with the FRB's central engine, likely through synchrotron emission.

Before the discovery of PRSs, theoretical studies had predicted steady synchrotron radiation from FRB central engines \citep{Yang2016, Murase2016}. Following the detection of the PRS linked to FRB 20121102A, multiple models emerged to explain PRS origins \citep{Dai2017, Metzger2017, Margalit2018, Margalit2018b, Margalit2019, Li2020c, Zhao2021, Sridhar2022}. In particular, \citet{Yang2020} proposed a generic model of synchrotron-emitting nebula and derived a simple linear relation between PRS luminosity and FRB rotation measure (RM). This “Yang relation” (Equation~\ref{eq:yang-zhang}) governed by the magneto-ionic environment of FRBs, naturally explains why FRB 20121102A (with the highest known RM) hosts a bright PRS, while FRBs with lower RM values typically lack such counterparts. Recent detections of three additional PRSs \citep{Niu2022, Bruni2024, Zhang2024c, Bhusare2024, Bruni2025, Zhang2025} and two candidate systems \citep{Ibik2024} further validate the Yang relation, with no significant outliers observed to date.  

The Yang relation enables intrinsic luminosity estimation through RM observations, establishing PRSs as viable cosmological probes. We apply this relation to constrain \(H_0\). Given that the number of confirmed PRS systems remains limited, our analysis focuses on a mock PRS sample generated through Monte Carlo simulations, which illustrates the potential of applying the Yang relation as a standard candle while also providing preliminary constraints on \(H_0\) by observed PRSs.

The remainder of this paper is organized as follows: Section~\ref{sec:method} provides a brief description of the Yang relation and the setup of the PRS population for Monte Carlo simulations. Section~\ref{sec:result} outlines the analysis procedures and results. Finally, Section~\ref{sec:discuss} summarizes our findings and discusses possible improvements to the methodology.

\end{CJK*}
\section{Method} \label{sec:method}
\subsection{Luminosity, Rotation Measure, and Dispersion Measure}
The luminosity of PRSs is related to the RM via the relation presented by \citet{Yang2020, Yang2022, Bruni2024}:
\begin{equation} \label{eq:yang-zhang}
L_{\nu} = \frac{64\pi^{3}}{27}m_{e}c^{2}\zeta_{e}\gamma_{\rm th}^{2}R^{2}\bigl|\mathrm{RM}_{\mathrm{src}}\bigr|,
\end{equation}
where \(\mathrm{RM}_{\mathrm{src}} = (1+z)^{2}\mathrm{RM}_{\mathrm{obs}}\)\footnote{In principle, the observed RM consists of 5 components like DM in 
Equation~\ref{eq:total_DM}. 
However, due to the weak and random IGM magnetic field, RM from IGM could be ignored.
From observation and simulation RM map of the Milky Way \citep{Reissl2023}, the Milky Way RM can also be neglected, even for low galactic latitude FRBs (FRB 20121102A and FRB 20201124A).
This issue should be carefully considered in the future when there are more PRS sources.}. Here, \(R\) denotes the emission region radius, \(\zeta_e\)  the ratio between the relativistic and nonrelativistic electron numbers, and \(\gamma_{\rm th}\) the minimum Lorentz factor of non-thermal electrons.  In this picture, the FRB central engine is at the center of a synchrotron nebula. The  relativistic electrons primarily dominate the synchrotron luminosity, while thermal electrons dominate the RM, because \(\mathrm{RM} \propto (\gamma ^2 m_e)^{-1}\)\citep{Quataert2000,Yang2022}. 
The characteristic Lorentz factor \(\gamma_{\rm th}\) marks the transition between these two populations, representing the typical energy that separates the thermal electrons responsible for RM from the more energetic electrons that drive the observed  emission.

Combining the luminosity relation with standard formula for flux density \(L_{\nu}  = 4\pi F_{\nu} \frac{1}{1+z} [\frac{c (1+z)}{H_{0}} \int_{0}^{z} \frac{1}{E(z')} \mathrm{d}z']^2\) leads to:
\begin{equation} \label{eq:RMobs}  
\begin{split}  
|\mathrm{RM}_{\mathrm{obs}}|  
&= \frac{27}{16\pi^{2} m_{e}} \frac{1}{\zeta_e \gamma_{\mathrm{th}}^{2} R^{2} H_{0}^{2}} \\  
&\quad \times \frac{F_{\nu}}{ 1+z} \left( \int_{0}^{z} \frac{1}{E(z')} \mathrm{d}z' \right)^{\!2}, 
\end{split}  
\end{equation}
where  \(E(z) = H(z)/H_0\) is the Hubble parameter normalized by the Hubble constant. We focus on the \(\Lambda\)CDM framework with \(E(z)=\sqrt{\Omega_m (1+z)^3 + 1 - \Omega_m}\) in this work.
Notice the degeneracy between the product \(\zeta_e \gamma_{\mathrm{th}}^{2} R^{2}\) and the Hubble constant \(H_0\).
Recent observational findings suggest that the product \(\zeta_e \gamma_{\mathrm{th}}^{2} (R / 0.01~\mathrm{pc})^{2}\) is approximately 1 \citep{Bruni2025}. However, simply calculating the luminosity involves \(H_0\), creating a circular argument.
To avoid this issue and break the degeneracy, we need an independent constraint for \(\zeta_e \gamma_{\mathrm{th}}^{2} R^{2}\). One natural method for obtaining this constraint is by using DM.

The total observed DM is the sum of contributions from multiple astrophysical components \citep[e.g.,][]{Thornton2013, Deng2014}:
\begin{equation} \label{eq:total_DM}
\mathrm{DM}_{\mathrm{obs}} 
= \mathrm{DM}_{\mathrm{MW}} 
+ \mathrm{DM}_{\mathrm{halo}} 
+ \mathrm{DM}_{\mathrm{IGM}} 
+ \frac{ \mathrm{DM}_{\mathrm{host}} + \mathrm{DM}_{\mathrm{src}} }{1+z},
\end{equation}
where \(\mathrm{DM}_{\mathrm{MW}}\), \(\mathrm{DM}_{\mathrm{halo}}\), \(\mathrm{DM}_{\mathrm{IGM}}\), \(\mathrm{DM}_{\mathrm{host}}\), and \(\mathrm{DM}_{\mathrm{src}}\) denote the contributions from the Milky Way's interstellar medium, Galactic halo, intergalactic medium (IGM), host galaxy's interstellar medium, and the FRB's local environment, respectively. The Galactic contribution, \(\mathrm{DM}_{\mathrm{MW}}\), is estimated using the YWM16 \citep{YMW16} or NE2001 \citep{NE2001} electron density models. 
For the present analysis, we focus mainly on the extragalactic contributions:
\begin{equation} \label{eq:DMex}
\begin{split} 
\mathrm{DM}_{\mathrm{ex}} 
&= \mathrm{DM}_{\mathrm{obs}} - \mathrm{DM}_{\mathrm{MW}} - \mathrm{DM}_{\mathrm{halo}} \\
&= \mathrm{DM}_{\mathrm{IGM}} +\frac{ \mathrm{DM}_{\mathrm{host}} + \mathrm{DM}_{\mathrm{src}} }{1+z}.
\end{split}
\end{equation}

The average IGM DM is expressed as \citep[e.g.,][]{Deng2014, Macquart2020}:
\begin{equation}\label{eq:DMigm}
\mathrm{DM}_{\mathrm{IGM}}(z)
= \frac{3cf_{\mathrm{IGM}}\Omega_bH_0^2}{8\pi G m_p H_0}
\int_{0}^{z} 
\frac{\chi(1+z')}{E(z')}
\mathrm{d}z',
\end{equation}
where \(\chi = 7/8\) denotes the electron fraction from ionized hydrogen and helium. Here we adopt \(f_{\mathrm{IGM}} \approx 0.83\) following \citep{Fukugita1998}, but again note the small difference between $f_{\mathrm{IGM}}$ and $f_{\mathrm{diff}}$ (up to 5\%, e.g. from the simulation of \cite{zhang2025b}).  The degeneracy between \(\Omega_b h^2\) and \(H_0\) is alleviated by adopting \(\Omega_b h^2 \approx 0.0223\) \citep{Wu2022}, as derived from quasar absorption observations \citep{Cooke2018}.

Previous literature frequently adopt \(\mathrm{DM}_{\mathrm{host}} + \mathrm{DM}_{\mathrm{src}} \sim 100\ \mathrm{pc\ cm^{-3}}\) to extract \(\mathrm{DM}_{\mathrm{IGM}}\) for cosmological studies \citep[e.g.][]{Li2020,Zhang2023}. 
This simple treatment, however, may not apply to our analysis, which focuses on the PRS-associated FRBs that typically have larger values of \(\mathrm{DM}_{\mathrm{host}} + \mathrm{DM}_{\mathrm{src}}\) 
\citep[e.g.][]{Yang2017b, Niu2022, Ravi2021, Chen2025}. This is consistent with the implication from their magneto-ionic environments \citep{Michilli2018Nature, Xu2022,Anna-Thomas2023}. Given these uncertainties, we treat \(\mathrm{DM}_{\mathrm{host}} + \mathrm{DM}_{\mathrm{src}}\) as a free parameter in our analysis, avoiding biased assumptions.

By eliminating \(H_0\) from Equation~\ref{eq:DMigm} using Equation~\ref{eq:RMobs} and substituting the result into Equation~\ref{eq:DMex}, we derive:
\begin{equation} \label{eq:DMexObs}
\begin{split}  
\mathrm{DM}_{\mathrm{ex}}  
&= \left( \frac{m_{e}c^{2}f_{\mathrm{IGM}}^{2}\chi^2\Omega_{b}^{2}H_{0}^{4}}{12G^{2}m_{p}^{2}}  \right. \\  
&\left. \quad \times \frac{\zeta_e\gamma_{c}^{2}R^{2}(1+z)|{\rm RM_{{\rm obs}}}|}{F_{\nu}}\right)^{1/2}  \frac{\int_{0}^{z}\frac{1+z'}{E(z')}\mathrm{d}z'}{\int_{0}^{z}\frac{1}{E(z')}\mathrm{d}z'}\\
&\quad  + \frac{ \mathrm{DM}_{\mathrm{host}} + \mathrm{DM}_{\mathrm{src}}}{1+z}. 
\end{split}  
\end{equation}
We can constrain \(\zeta_e \gamma_{\mathrm{th}}^{2} R^{2}\) by fitting \(\mathrm{DM}_{\mathrm{ex}}\) with \(z, F_{\nu}\) and \(\rm RM_{{\rm obs}}\).

The uncertainty of \(\mathrm{DM}_{\mathrm{ex}} \) is quantified as
\begin{equation} \label{eq:sigma_dm}
\sigma_{ { \mathrm{DM} }_{ \mathrm{ex} }}
= \sqrt{ \sigma_{\mathrm{MW}}^2 + \sigma_{\mathrm{IGM}}(z)^2 + \sigma_{\mathrm{host+src}}^2} ,
\end{equation}
where \(\sigma_{\mathrm{MW}} \approx 30~\mathrm{pc~cm^{-3}}\) (from the uncertainty of Galactic electron number density model \citep{Manchester2005}) represents the combined uncertainty from \(\mathrm{DM}_{\mathrm{MW}}\) and \(\mathrm{DM}_{\mathrm{halo}}\). 
The term \(\sigma_{\mathrm{IGM}}(z)\) denotes the uncertainty of \(\mathrm{DM}_{\mathrm{IGM}}\).
Following \citet{Li2020}, we derive it by fitting numerical simulations from \citet{McQuinn2014}. 
We adopt $\sigma_{\mathrm{host+src}} \approx 100~\mathrm{pc~cm^{-3}}$, a conservative estimate that exceeds the typical values of the FRB observations \citep{Chittidi2021,Bernales--Cortes2025} and simulations \citep{Zhang2020b,Mo2023,Theis2024,Kovacs2024,Orr2024,Zhang2025}, to account for additional environmental uncertainties in PRS systems.

\begin{figure}[htb!]
    \centering
    \includegraphics[width=1\linewidth]{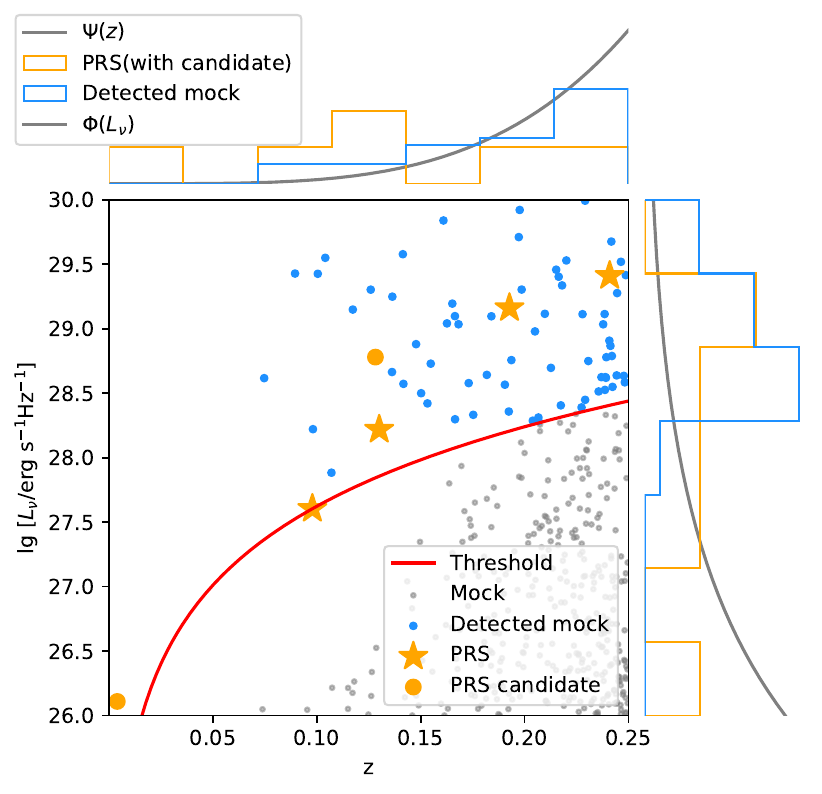}
    \caption{Observed and simulated PRS population. 
    \textit{Main panel}: Luminosity-redshift distribution for mock (grey/blue) and observed PRSs (orange). The red curve indicates the detection threshold. \textit{Upper/right panels}: Intrinsic redshift distribution \(\Psi(z)\) and luminosity function \(\Phi(L_\nu)\) (grey), compared to detected mock PRSs (blue).}
    \label{fig:MC}
\end{figure}

\subsection{PRS redshift and luminosity distribution} \label{sec:MC}

To validate our methodology, we employ Monte Carlo simulations to generate mock PRS samples and reconstruct $H_0$. This process requires two population functions: a redshift distribution and a luminosity function for PRSs. Although our analysis is not critically dependent on the true forms of these functions, we adopt observationally motivated models to replicate observed trends and provide preliminary predictions for the PRS population.

The origin of FRBs remains under debate. Two possibilities have been suggested and gained observational support. One hypothesis proposes that they arise from core-collapse supernovae, implying that the redshift distribution follows the cosmic star formation history \citep[e.g.,][]{James2022, James2022b, Shin2023, Wang2024}. Alternatively, there are suggestions that at least a fraction of FRBs have an origin with a delay with respect to star formation history\citep[e.g.,][]{Zhang2022, zzl2023, Gupta2025}. 

For the specific sample we are studying, current data strongly associate PRS-linked FRBs with young stellar environments, since every confirmed PRS-FRB system has been observed in a dwarf galaxy or an actively star-forming galaxy \citep{Tendulkar2017, Li2019, Li2020b, Niu2022, Piro2021, Ravi2021, Chen2025}. Even the two candidate PRSs appear to be hosted by star-forming galaxies \citep{Ibik2024}. Consequently, we adopt the cosmic star formation history as the basis for the redshift distribution \(\Psi(z)\) of PRS-associated FRBs, using the Equation~(15) of \citet{Madau2014} (shown as the grey solid line in the upper panel of Figure~\ref{fig:MC}).\footnote{A conversion from event-rate density to redshift rate distribution is needed \citep[e.g.,][]{Zhang2021}.} In addition, we assume a power-law luminosity function of the form \(\Phi(L_\nu) \propto L_\nu^{-1.3}\) (also illustrated by the grey solid line in the upper panel of Figure~\ref{fig:MC}) for PRSs, which closely reproduces the observed luminosity distribution.

\section{Result} \label{sec:result}

We perform Markov Chain Monte Carlo (MCMC) analyses to estimate \(H_0\) using about 50 mock PRS detections and observational data.
The observational sample includes 4 confirmed PRSs \citep{Chatterjee2017, Niu2022, Bruni2024, Zhang2024c, Bhusare2024, Bruni2025, Zhang2025} and 2 candidate PRSs \citep{Ibik2024}.  

\begin{figure}[htb!]
    \centering
    \includegraphics[width=1\linewidth]{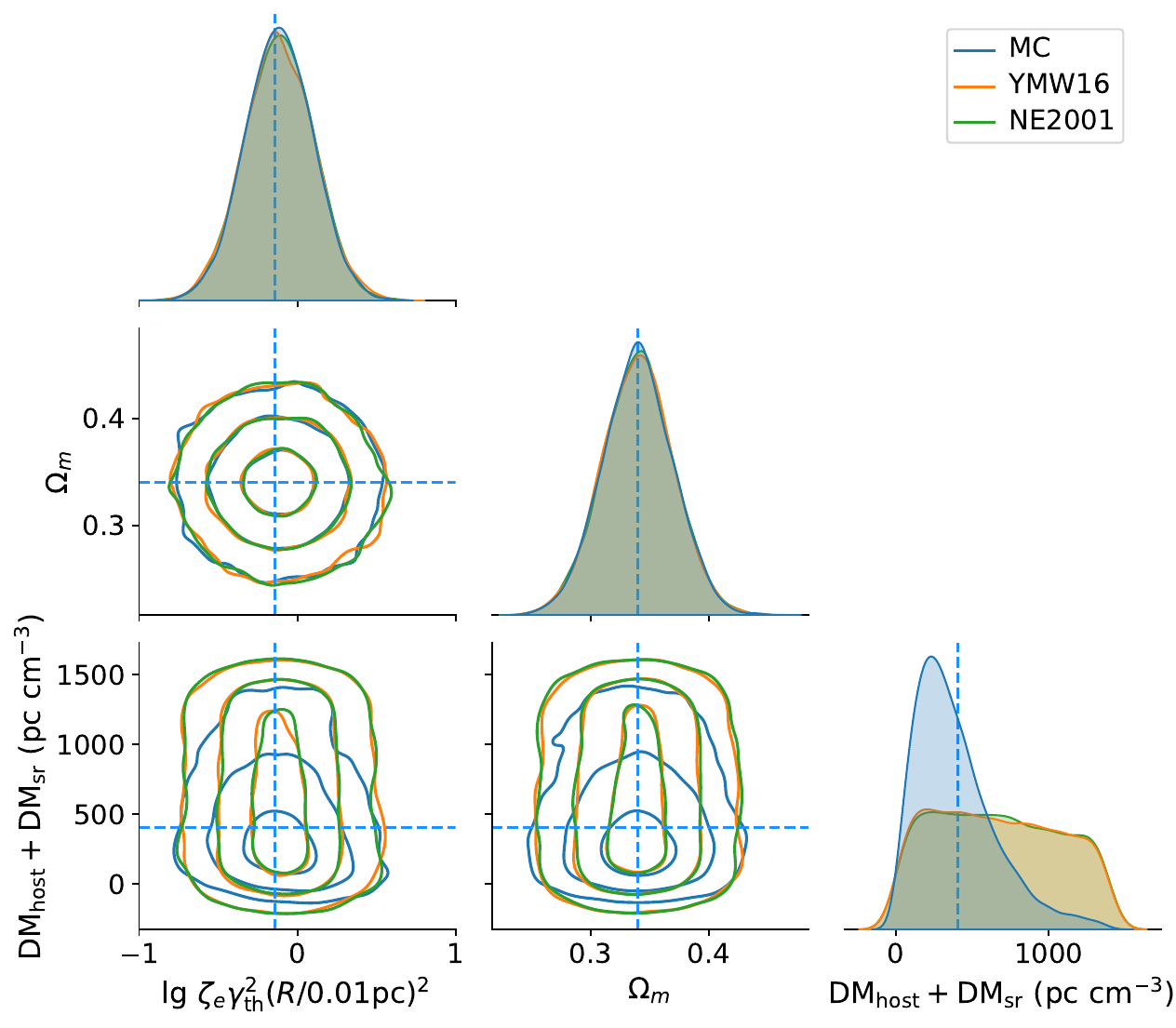}
    \caption{Posterior distributions from MCMC analyses of Equation~\ref{eq:DMexObs}.  
    The contour plot is smoothed using kernel density estimation. Blue, orange, and green solid lines correspond to fitting results for detected mock PRSs and real PRSs analyzed using the YMW16 model and the NE2001 model, respectively.  
    In each contour set, 1$\sigma$, 2$\sigma$, and 3$\sigma$ confidence regions appear from innermost to outermost.  
    Blue dashed lines indicate the median values of the injected distributions of the Monte Carlo simulation.
    }
    \label{fig:RMDM2}
\end{figure}

\begin{figure}[htb!]
    \centering
    \includegraphics[width=1\linewidth]{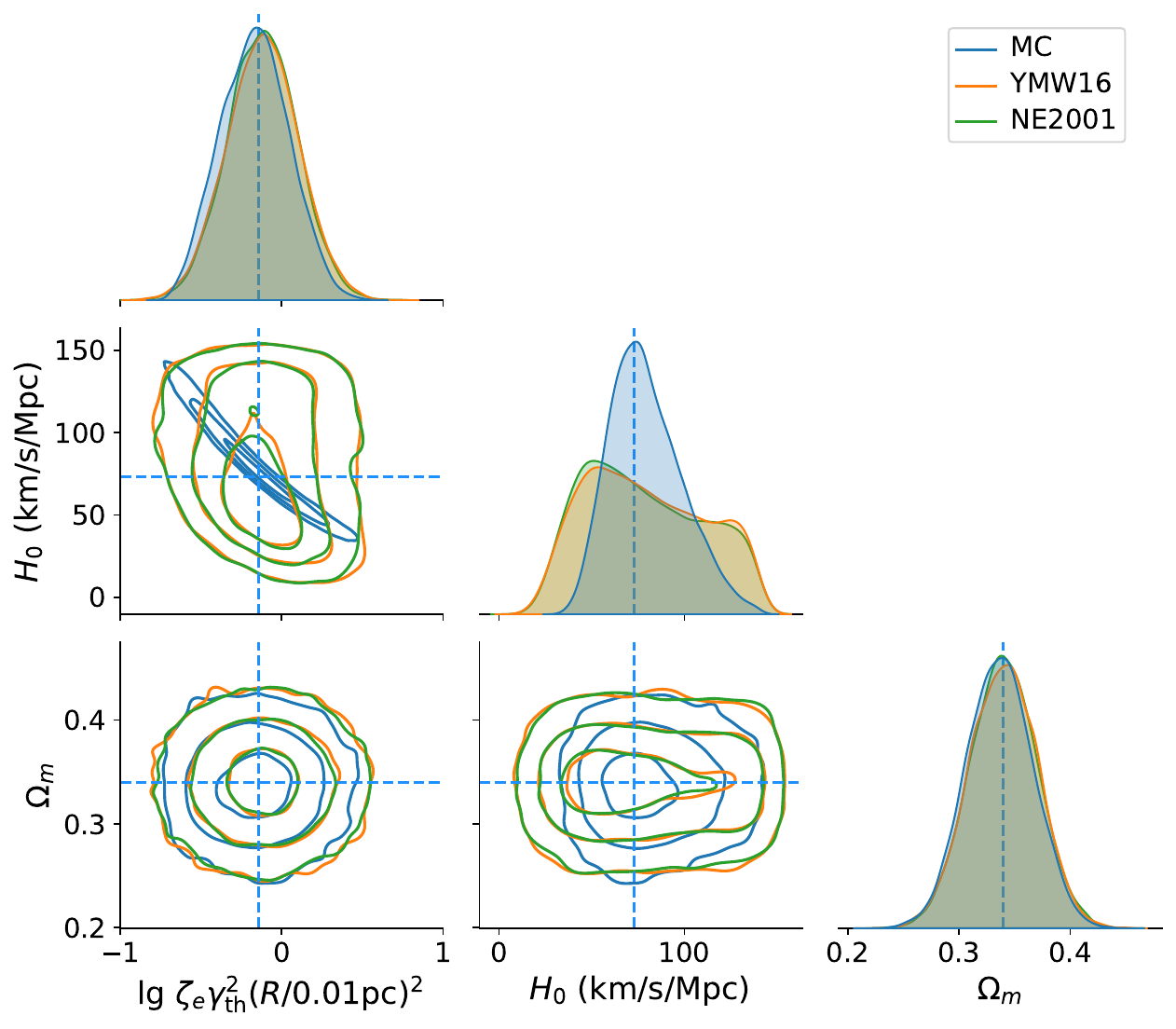}
    \caption{Posterior distributions from MCMC analyses of Equation~\ref{eq:RMobs}. The posterior distribution of \(\mathrm{lg}\  \zeta_e \gamma_{\mathrm{th}}^{2} (R / 0.01~\mathrm{pc})^{2}\) in Figure~\ref{fig:RMDM2}  is adopted as the prior distribution.}
    \label{fig:RM_fit}
\end{figure}

The analysis proceeds as follows:  
\begin{enumerate}
    \item \textbf{Mock Sample Generation}: Mock PRS sources are generated by drawing \((z, L_\nu)\) pairs from the redshift distribution \(\Psi(z)\) and luminosity function \(\Phi(L_\nu)\), shown as grey and blue points in Figure~\ref{fig:MC}. The specific flux \(F_\nu\) for each source is calculated, and sources exceeding the detection threshold of \(0.02~\mathrm{mJy}\) (the minimum sensitivity for detected PRSs) are classified as detectable (blue points in Figure~\ref{fig:MC}). 
    
    \item \textbf{Parameter Initialization}: For the detectable mock PRSs, \(\mathrm{DM}_{\mathrm{IGM}}\) is computed using Equation~\ref{eq:DMigm} with \(\Omega_m = 0.34\) and \(H_0 = 73.04~\mathrm{km~s^{-1}~Mpc^{-1}}\) \citep{Riess2022}, since FRBs are local universe phenomena. The parameter \(\zeta_e \gamma_{\mathrm{th}}^{2} (R / 0.01~\mathrm{pc})^{2}\) is assigned values from a log-normal distribution (\(\mu = -0.333\), \(\sigma = 0.6\)) \citep[e.g.,][]{Bruni2025}, and \(|\mathrm{RM}_{\mathrm{src}}|\) is calculated using Equation~\ref{eq:yang-zhang}. The combined \(\mathrm{DM}_{\mathrm{host}} + \mathrm{DM}_{\mathrm{src}}\) is drawn from a log-normal distribution (\(\mu = 6\), \(\sigma = 0.6\)). Finally, \(\mathrm{DM}_{\mathrm{ex}}\) values for mock PRS are derived using Equation~\ref{eq:DMexObs}. 
    
    \item \textbf{First-Stage MCMC}: Equation~\ref{eq:DMexObs} is fitted to each sample set, with the uncertainty shown in Equation~\ref{eq:sigma_dm}. Priors include a Gaussian distribution for \(\Omega_m = 0.34 \pm 0.03\) to include both early- and late-universe result \citep{Planck2018, Riess2022}, a log-normal distribution for \( \zeta_e \gamma_{\mathrm{th}}^{2} (R / 0.01~\mathrm{pc})^{2}\) with (\(\mu=0\), \(\sigma=0.5\)), and a uniform prior for \(\mathrm{DM}_{\mathrm{host}} + \mathrm{DM}_{\mathrm{src}}\) within \((0, 1500)~\mathrm{pc~cm^{-3}}\). Results are shown in Figure~\ref{fig:RMDM2}.  
    
    \item \textbf{Second-Stage MCMC}: The posterior distribution of \(\zeta_e \gamma_{\mathrm{th}}^{2} (R / 0.01~\mathrm{pc})^{2}\) from the first stage is used as a prior to fit Equation~\ref{eq:RMobs}. A uniform prior for \(H_0\) is adopted within \((0, 140)\).  Results are shown in Figure~\ref{fig:RM_fit}. The final posterior distribution of \(\zeta_e \gamma_{\mathrm{th}}^{2} (R / 0.01~\mathrm{pc})^{2} \sim 1\) is expected as the first glance at luminosity-RM relation \citep{Bruni2025}.
    We get \(H_0 = 75 \pm 30 ~\mathrm{km~s^{-1}~Mpc^{-1}}\) for observed PRSs by YMW16 and NE2001 model, which shows our method is insensitive to the Galactic electron number density models. For the mock sample, we get \(H_0 = 75 \pm 15 ~\mathrm{km~s^{-1}~Mpc^{-1}}\), consistent with our injection assumption.
\end{enumerate}

\section{Summary and Discussion} \label{sec:discuss}
\begin{itemize}  
  \item Traditional FRB cosmology struggles with overlapping dependencies among \(H_0\), \(\Omega_b\), and \(f_{\mathrm{IGM}}\). The Yang relation \citep{Yang2020} can help to disengage this by incorporating \(\mathrm{RM}\) to link PRS luminosity to the magneto-ionic environment of FRBs, alleviating degeneracies inherent in DM-only methods.

  \item The principal aim of this study is to demonstrate the feasibility of using the Yang relation to constrain cosmological parameters. We begin by generating a mock PRS sample. As a secondary investigation, we perform a preliminary population study of PRSs. We find that the redshift distribution follows cosmic star formation, the luminosity function follows a power law  (\(\Phi(L_\nu) \propto L_\nu^{-1.3}\)) and the parameter \(\zeta_e\gamma_{\mathrm{th}}^{2}(R/0.01\,\mathrm{pc})^{2}\) follows a log-normal distribution (\(\mu = -0.333\), \(\sigma = 0.6\)), in line with current observational data.  

  \item We introduce a two-stage MCMC framework to constrain \(H_0\). First, we derive the product \(\zeta_e\gamma_{\mathrm{th}}^{2}(R/0.01\,\mathrm{pc})^{2}\) using DM observations (Equation~\ref{eq:DMexObs}), then use the result as priors to estimate \(H_0\) (Equation~\ref{eq:RMobs}). For our mock sample, this procedure yields \(H_0 = 75 \pm 15 ~\mathrm{km~s^{-1}~Mpc^{-1}}\), consistent with the injected value. Applying the method to six observed PRSs gives \(H_0 = 75 \pm 30 ~\mathrm{km~s^{-1}~Mpc^{-1}}\) for both YMW16 and NE2001 model, indicating insensitivity to specific Galactic density models.

  \item However, a degeneracy remains between \(\zeta_e\gamma_{\mathrm{th}}^{2}R^2\) and \(H_0\) (Figure~\ref{fig:RM_fit}), highlighting the need for further refinements. Future advancements will benefit from a larger PRS sample enabled by interferometric facilities (e.g., CHIME Outrigger, \citealt{Lanman2024}; MeerKAT, \citealt{Jonas2016}; ASKAP, \citealt{Hotan2021}; VLA, \citealt{Perley2011}; DSA-2000, \citealt{2024AAS...24323705H}; and SKA, \citealt{Dewdney2009}), which improve the statistics. With a bigger sample, a full Bayesian treatment, incorporating carefully considering prior distributions of all parameters (e.g., different DM components, \(f_{\mathrm{IGM}}\), \(\Omega_b\)),  can refine constraints on cosmological parameters \citep[e.g.,][]{Macquart2020,Wu2022}.
  
  \item Improved constraints on \(\zeta_e\gamma_{\mathrm{th}}^{2}R^2\) may be obtained through observation and theoretical modeling.
  Observationally, high-resolution radio imaging and multiwavelength observations can constraint the physical size \(R\) of the nebula and thereby limit \( \zeta_e \gamma_{th}\). For example, recent observations of the hyperactive FRB~20240114A show an evolving absorption frequency in its PRS \citep{Zhang2025}, which could be related to \(\gamma_{\rm th}\). Theoretically, modeling PRSs within frameworks such as supernova remnant or magnetar wind nebula models can further clarify their evolutionary patterns.

  \item A promising strategy for using the Yang relation is to calibrate \(\zeta_e\,\gamma_{\mathrm{th}}^{2}R^2\) by observing another standard candle in the same host galaxy, akin to the approach of calibrating absolute magnitudes for Type Ia supernovae \citep[e.g.,][]{Kowal1968, Branch1992}. 
  This approach avoids uncertainties and degeneracies inherent to pure DM methods. The persistent nature of PRSs enables repeated host galaxy observations targeting Type~Ia supernovae, which provide independent constraints on \(\zeta_e\gamma_{\mathrm{th}}^{2}R^{2}\). Assuming the luminosity of FRB~20190520B's PRS, we estimate the detectability of PRSs out to \(z \sim 0.8\), which demonstrates PRSs as viable cosmological probes.

  \item In principle, this strategy can be applied to any persistent radio source exhibiting analogous characteristics, regardless of its current association with FRBs. \citet{Dong2024b} have identified continuous compact radio sources with VLA that are not linked to FRBs but may represent older PRSs; if these sources share the same emission mechanism as those associated with FRBs and have \(\mathrm{RM}\) measured\footnote{Notice that the RM of FRBs can be measured from its bright, polarized pulses, it remains uncertain whether the continuous emission from a persistent source is sufficiently polarized to allow a robust RM measurement.}, they could similarly be used to constrain cosmological parameters using the Yang relation.

\end{itemize}

\begin{acknowledgements}

Z.Z. recalls that this idea first came to his mind when his thoughts drifted off during an astrophysics class.
Z.Z. thanks Hua Chen and Jia-Ming Zhu-Ge for helpful discussions. We thank the referee for helpful comments. This work is supported by the Nevada Center for Astrophysics, a TTDGRA fellowship at UNLV, and NASA 80NSSC23M0104.
\end{acknowledgements}


\software{astropy \citep{Astropy2022}, NumPy \citep{NumPy}, Matplotlib \citep{Matplotlib}, emcee \citep{emcee}, SciPy \citep{SciPy}, seaborn \citep{seaborn}}

\newpage
\bibliography{references}{}
\bibliographystyle{aasjournal}



\end{document}